\documentclass[prb,aps, longbibliography, preprintnumbers,11pt]{revtex4-1}
\usepackage{amssymb,bm,setspace,enumerate,amsmath}
\usepackage{cases}
\usepackage{graphicx}
\usepackage[bookmarks, colorlinks=true, breaklinks]{hyperref}  
\hypersetup{linkcolor=blue,citecolor=blue,filecolor=black,urlcolor=blue} 
\usepackage{color, soul,ulem}
\usepackage{caption}
\usepackage{adjustbox}

\renewcommand{\vec}[1]{\mathbf{#1}}
\newcommand{\sv}[1]{_{\vec{#1}}}
\newcommand{\svk}{_{\vec{k}}}

\newcommand{\svr}{\sv{r}}

\newcommand{\svdispx}{_{\vec{r}-\vec n_x}}
\newcommand{\svdispy}{_{\vec{r}-\vec n_y}}

\newcommand{\svpik}{_{\vec k-\vec{M}}}

\newcommand{\tx}{t_{k_x}}
\newcommand{\ty}{t_{k_y}}
\newcommand{\tplus}{(t_{k_x}^2+t_{k_y}^2)}
\newcommand{\tmin}{(t_{k_x}^2-t_{k_y}^2)}
\newcommand{\tprod}{t_{k_x} t_{k_y}}
\newcommand{\tprodsq}{t_{k_x}^2 t_{k_y}^2}
\newcommand{\tminfour}{(t_{k_x}^4- t_{k_y}^4)}

\newcommand{\bra}{\left< }
\newcommand{\ket}{\right>}

\makeatletter 
\renewcommand{\fnum@figure}{\textbf{Figure~\thefigure}}
\makeatother

\begin{document}

\title{Cooperative elastic fluctuations provide tuning of the metal-insulator transition}

\author{G. G. Guzm\'{a}n-Verri$^{1,2,3}$\footnote{gian.guzman@ucr.ac.cr}, R. T. Brierley$^4$, P. B. Littlewood$^{3,5}$\footnote{pblittlewood@anl.gov}}
\affiliation{$^{1}$Centro de Investigaci\'{o}n en Ciencia e Ingenier\'{i}a de Materiales~(CICIMA), Universidad de Costa Rica, San Jos\'{e}, Costa Rica 11501,}
\affiliation{$^{2}$Escuela de F\'{i}sica, Universidad de Costa Rica, San Jos\'{e}, Costa Rica 11501,}
\affiliation{$^{3}$Materials Science Division, Argonne National Laboratory, Argonne, Illinois, USA 60439,}
\affiliation{$^{4}$Department of Physics, Yale University, New Haven, Connecticut 06511, USA,}
\affiliation{$^{5}$James Franck Institute, University of Chicago, 929 E 57 St, Chicago, Illinois, USA 60637.}

\date{\today}
\maketitle

 \textbf{Metal to insulator transitions~\cite{Imada1998a} (MITs) driven by strong electronic correlations are common in condensed matter systems, and are associated with some of the most remarkable collective phenomena in solids, including superconductivity and magnetism. Tuning and control of the transition holds 
the promise of novel, low power, ultrafast electronics~\cite{Yang2011a}, but the relative roles of doping, chemistry, elastic strain and other applied fields has made systematic understanding difficult to obtain. Here we point out that existing data~\cite{Torrance1992a, Hwang1995a, RodriguezMartinez1996a} on tuning of the MIT in perovskite transition metal oxides through ionic size effects provides evidence of systematic and large effects on the phase transition due to dynamical fluctuations of the elastic strain, which have been usually neglected~\cite{Khomskii2014a}. This is illustrated by a simple yet quantitative statistical mechanical calculation in a model that incorporates cooperative lattice distortions coupled to the electronic degrees of freedom.
We reproduce the observed dependence of the transition temperature on cation radius in the well-studied manganite~\cite{Tokura2006a} and nickelate~\cite{Catalano2018a} materials. Since the elastic couplings are generically quite strong, these conclusions will broadly generalize to all MIT’s that couple to a change in lattice symmetry.}

MITs driven by electronic correlations have energy scales in the electron volts, yet it is common to find that these phase transitions happen at temperatures corresponding to much lower energies~\cite{Imada1998a}.  In the absence of a mechanism of fine tuning the coupling constants, it is natural to look for entropic rather than enthalpic contributions to describe these transitions.
Since all observed MITs couple to the lattice, one is then driven to look for phononic entropic contributions. As a hint to the origin of these interactions,
\begin{figure}[htp]
  \centering
   \includegraphics[scale=0.6]{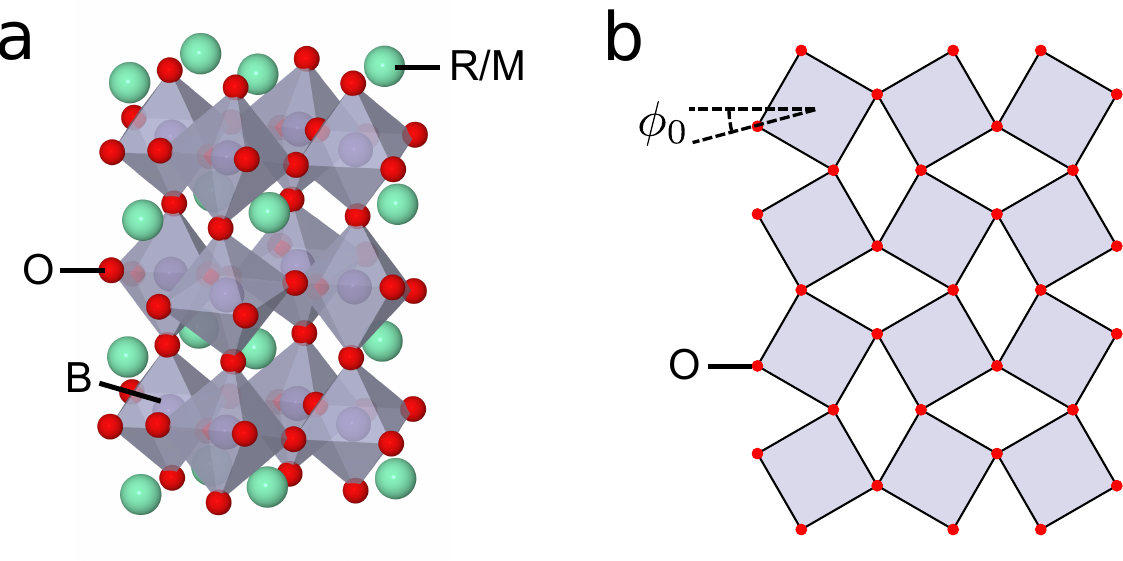}
  \caption{ {\bf Pervoskite lattices.} {\bf (a)} 3D perovskite lattice showing the tilts of the BO$_6$ (B=Mn,Ni) octahedra.  R is a rare earch element such as  La, Pr, Nd, and Sm;  and M is an alkaline earth metal such as Ca, Sr, Ba. {\bf (b)} 2D representation of the tilts used in our model where $\phi_0$ is an initial equilibrium antiferrodistortive rotation.}
  \label{fig:2D_perovskite}
\end{figure}
a large number of transition metal oxides (TMO) with the ABO$_3$ perovskite crystal structure allow tuning of the MIT by not only by the choice and average valence of the electronically active B ion (usually a $3d$ transition metal) but also by the size of the electronically inactive A ion (usually a rare earth or alkaline earth)~\cite{Torrance1992a, Hwang1995a, RodriguezMartinez1996a, Katsufuji1997a}. This ‘size effect’ can shift the transition temperature $T_\text{MI}$ by hundreds of Kelvin, and the widely accepted explanation~\cite{Fujimori1992a} is that it is due to a reduction in the electron bandwidth as the bond bending induced by ionic size changes the orbital overlap.
However, the changes in bandwidth are not sufficiently large to explain such temperature variations~\cite{Sarma1994a, Radaelli1997a, Medarde1998a, Varignon2017a}. Moreover, it seems remarkable that a critical value of the ratio of interaction strength to bandwidth can be crossed in every $3d$ TMO, solely by varying the counterion~\cite{Fujimori1992a}.

Instead, we propose here that even when the transition is quite clearly driven by local electronic correlations, anisotropic long range forces induced by elastic compatibility conditions produce enormous entropic contributions to the free energy, which we show are crucial to describe the trends of the MIT with cation size. We illustrate this with a model of highly fluctuating cooperative lattice distortions that competes with a low temperature phase of constant free energy, i.e., a ferromagnetic metal (FM) for the manganites and a paramagnetic insulator (PMI) for the nickelates.  We do not aim to capture the complex charge, orbital, and magnetic orderings of these materials, but rather their high temperature melted version where the entropy is dominated by the cooperative distortions. Our view is that the natural experiments in the manganite and nickelate series broadly implicate elastic interactions as being important in a wide class of MITs, not only in the perovskites.

In building our model, we account for the electronic degrees of freedom by assuming we can separate the energy into components that can be calculated locally while keeping the long-range physics explicit. At zero Kelvin, state-of-the-art first-principles calculations can give such local free energy containing implicitly electron-phonon coupling on a unit cell as well as band structure energy and Coulomb correlation. We have not performed such calculations here. Instead, we have assumed that there is a simple functional outcome that can be parametrized, is the same across each material series, and is independent of the long-range piece.

Our approach has of course its limitations: not every TMO is electronically the same, e.g., the bandwidth is not the only indicator and/or key parameter of structural changes in the electronic structure when varying the rare-earth ion nor the local electronic correlations are independent of the tolerance factor \cite{Pavarini2005a, Han2018a}. These are idealizations which can only describe real materials approximately. Nonetheless, it allow us to illustrate that the non-trivial and surprisingly subtle effects from long range elastic interactions mediated between local degrees of freedom cannot be ignored when it comes to determine the structural trends of MITs that couple to symmetry breaking distortions.

The crystal structure of perovskite TMOs consists of corner-sharing oxygen octahedra surrounding the $B$ transition metal ion, as shown in Fig.~\ref{fig:2D_perovskite}\,(a). In general, the octahedra are tilted relative to their neighbors in an alternating pattern, and the tilt angle $\phi_0$ increases with smaller A-site cation radius $r_A$. 
The dramatic changes in the functional behavior of perovskites when varying $\phi_0$ have led to proposals~\cite{Rondinelli2012a} to engineer material properties by using a combination of strain, doping and pressure.
In addition to variations of the atomic size,  doping with A-site cations also introduces disorder in the cation size; careful distinction of the effects of doping and disorder for the manganites demonstrated that disorder reduces the $T_\text{MI}$ as effectively as varying $r_A$.~\cite{RodriguezMartinez1996a}

\begin{figure}[htp!]
  \centering
  \includegraphics[scale=0.5]{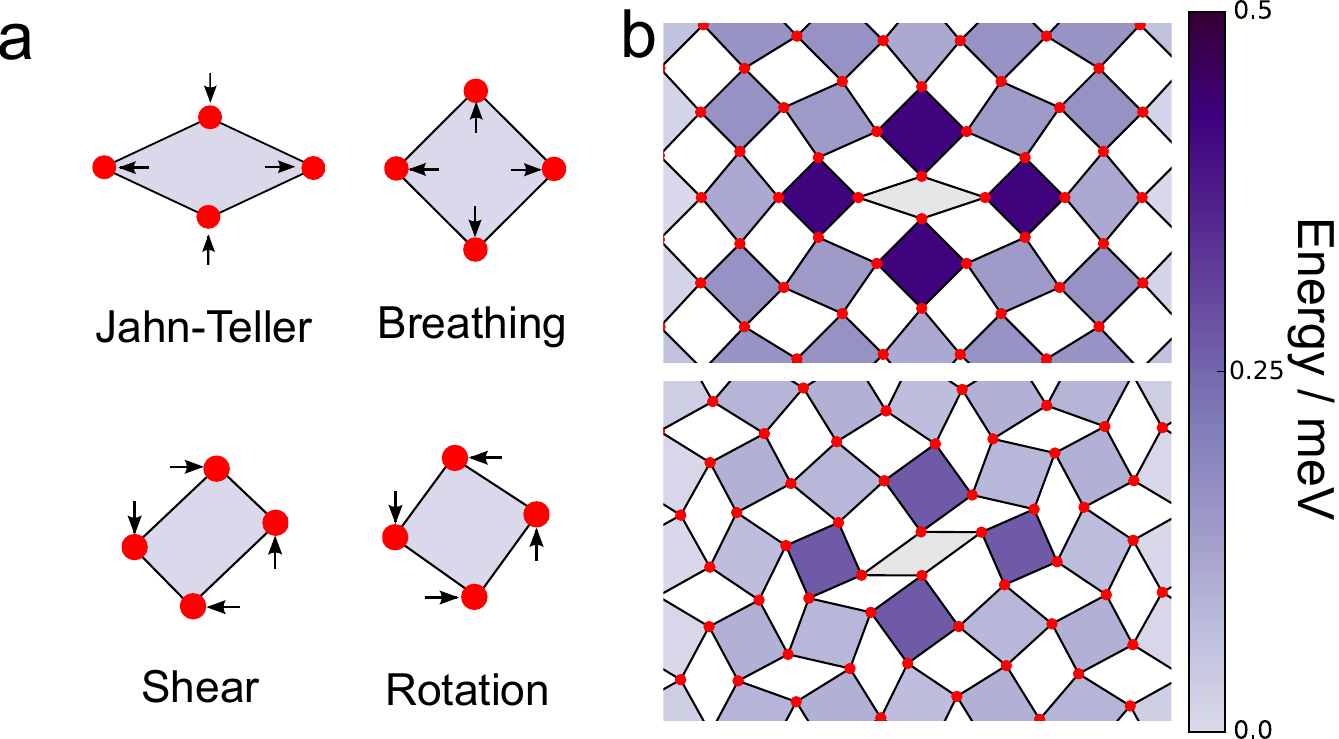}
  \caption{ {\bf Lattice distortions and strain responses.}  {\bf (a)} Lattice distortions considered in our model. {\bf (b)} 
  Strain responses of a lattice to a local JT distortion as a result of rotations. The color of each square indicates the strain energy associated with the local distortions of that square. The grey parallelogram at the centre has a JT distortion of fixed amplitude. The strain fields weaken by allowing the BO$_6$ to tilt, as the energy is more effectively absorbed locally. Additional distortions on this site, such as shear, are allowed. Top: lattice with $\phi_0=0$. Bottom: lattice with $\phi_0=15^\circ$.  }
  \label{fig:strain-response}
\end{figure}

Although purely electronic mechanisms to describe TMOs are appealing in their theoretical simplicity, it is known that the strong electron-phonon coupling means that the effects of lattice distortions cannot be neglected, and this is particularly well studied in manganites and nickelates~\cite{Millis1995a, Mercy2017a}. 
An electron that is localized by correlation effects in a unit cell will lower its energy further by the creation of a lattice distortion, which may be of different symmetry in different materials. In the nickelates this is a simple breathing distortion, and in the manganites a so-called Jahn-Teller~(JT) distortion that lowers the cubic symmetry of the octahedon,  as shown in Fig.~\ref{fig:strain-response}~(a).
The competition between this potential energy gain and the kinetic energy gained by delocalization to form a metal gives rise to the complex MIT phenomena in these materials. 

The corner-sharing constraint on the octahedra introduces compatibility conditions between distortions at different lattice sites; when integrating out the phonon degrees of freedom these yield highly anisotropic, long-range interactions~\cite{Kartha1995a}. Previous studies~\cite{Khomskii2014a, Ahn2004a, Ahn2013a} of phonon cooperativity in the manganites have demonstrated that they can explain the complex charge ordered phases and mesoscopic structures that have been observed in the manganites, and studied some effects of cooperative coupling on the transition~\cite{Millis1996a}. However, these studies did not consider the effect of octahedral tilting on the long-range interaction of the distortions.
The purpose of this work is to study such effects, and in doing so, to construct a complete theory for cooperative elastic effects at a phase transition.  

For illustration, we use a two-dimensional model of a  perovskite, where we replace the octahedra by squares, as shown in  Fig. 1(b).  Although the physics of bulk perovskites is three-dimensional, two-dimensional models~\cite{Ahn2004a, Ahn2013a} of elastic interactions capture their anisotropy and long-range decay (they fall-off as  $r^{-D}$ for $D=2$, and $3$ dimensions)
which in turn have been shown to generate structural inhomogeneity over a wide range of length scales. This is the most relevant aspect to our work and one of the most salient features 
that have been experimentally seen in TMOs.
At  a lattice site ${\vec r}$, the squares can undergo the  distortions shown in Fig.~\ref{fig:strain-response}\,(a):
deviatoric/JT modes $T\svr$,  dilatation/breathing modes  $D\svr$, shear modes $S\svr$, and small rotations $R\svr$ of the squares from an initial equilibrium antiferrodistortive rotation $\phi_0$, i.e., $\phi\svr = (-1)^{|\vec r|}\phi_0 + R\svr$.  
  Assuming a harmonic energy penalty for creating distortions from an equilibrium configuration,
\begin{gather}
  \label{eq:energy-unsubbed}
  H = \sum\svr a_T T\svr^2 + a_D D\svr^2 + a_S S\svr^2,
\end{gather}
combined with the corner-sharing constraint, we can find an effective interaction $V_{\vec r\vec r'}(\phi_0)$  between different types of distortion which gives rise to lattice cooperativity (see Supplementary Note 1).
$a_T$, $a_D$, and $a_S$ are, respectively, the stiffness of the JT, breathing, and shear distortions in a single, free octahedron and are independent of  $\vec{r}$. 

\begin{figure}[h!]
  \centering
     \includegraphics[scale=0.4]{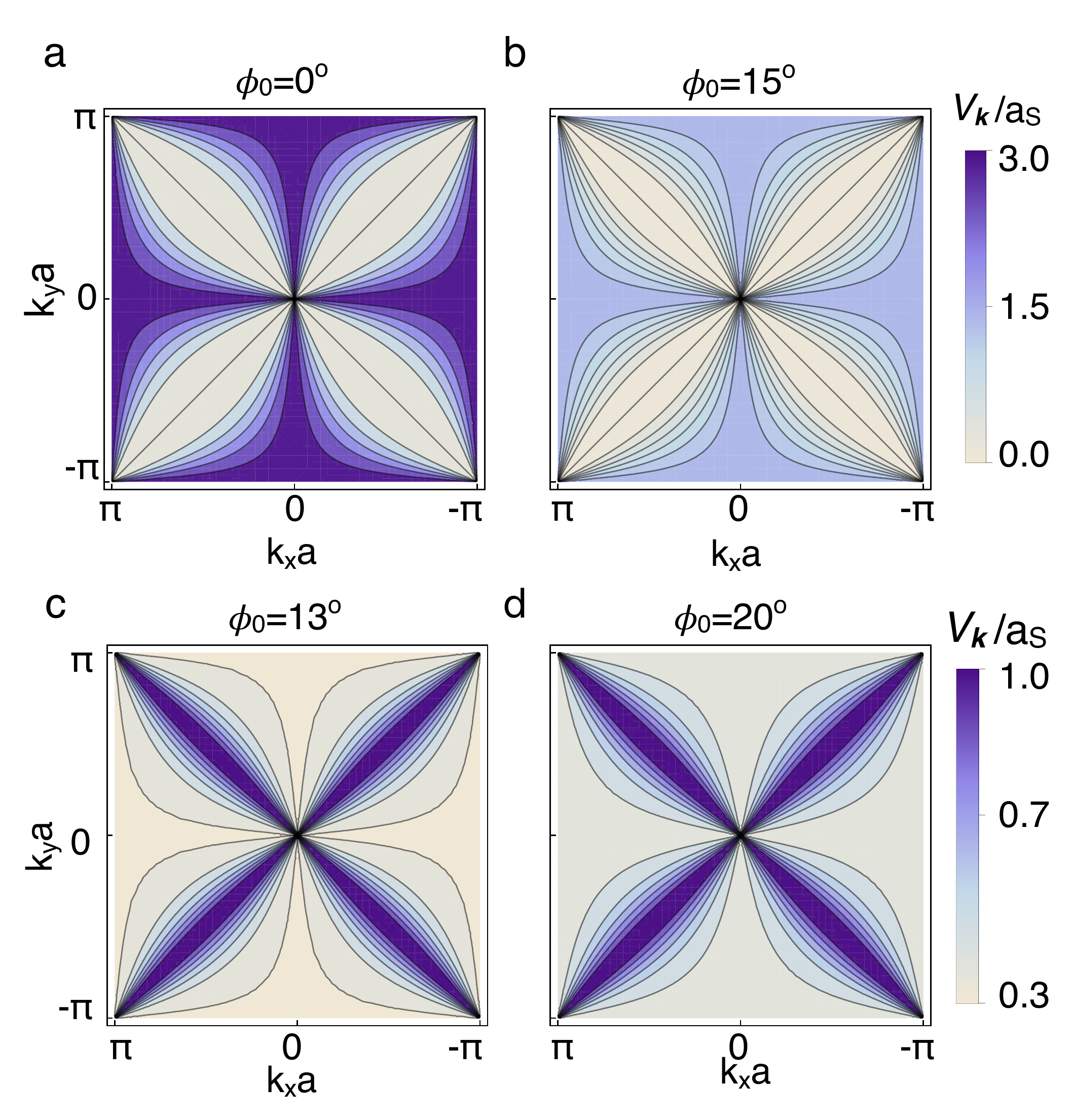}
  \caption{ {\bf Effective elastic energy.} Effective elastic  energy for JT distortions  in momentum space for {\bf (a)} $\phi_0=0$ and {\bf (b)} $\phi_0=15^\circ$. Rotations of the BO$_6$ octahedron allowed by the reduction of the A-cation size decrease the elastic energy. The characteristic butterfly pattern is a consequence of the anisotropy and long-range nature of strain forces, which in turn can generate the salient nano- and meso-scale structural inhomogeneities similar to those that have been observed in the manganites~\cite{Ahn2004a, Ahn2013a}  such as domain patterns in the form of stripes and tweeds formed by interwoven incommensurate structures.  Effective elastic  energy for breathing distortions  in momentum space for {\bf (c)} $\phi_0=0$ and {\bf (d)} $\phi_0=20^\circ$.  Tilts of the NiO$_6$ octahedron increase the effective elastic energy. The butterfly pattern is similar to that of the manganites which produces structural inhomogeneity.} 
  \label{fig:anisotropy}
\end{figure}

Fig.~\ref{fig:anisotropy} shows that the interaction strength is reduced by an increase tilt angle for JT distortions.
This occurs because in the tilted configuration it is possible for the distortion to be accomodated by additional rotations to the neighbouring sites, rather than changes in the shape.
Characteristic strain responses of the lattice to a local JT distortion with and without rotations are shown in Fig.~\ref{fig:strain-response}~(b).  

Both manganites~\cite{Tokura2006a} and nickelates~\cite{Catalano2018a} undergo first-order transitions from a characteristic low temperature phase to a high-temperature polaronic phase. This suggests that the motion of conduction electrons through the lattice is associated with the creation of local structural distortions that lead to bad metal behavior~\cite{Jaramillo2015a}. When the distortion interaction $V_{\vec r\vec r'}(\phi_0)$ is reduced by changes in $\phi_0$, the high-temperature phase is favoured by a reduction in the polaron formation energy~\cite{Millis1996a}. To study this behaviour, we  use $V_{\vec r\vec r'}(\phi_0)$  to form a statistical mechanical model for the distortions in this high-temperature phase, with a Hamiltonian,
\begin{align}
\label{eq:Hamiltonian}
 H &= \sum\svr \left[ \frac{1}{2} \Pi\svr^2 - \frac{\kappa}{2} Q\svr^2 + \frac{\gamma}{4} Q\svr^4 \right] + \sum_{\vec{r} \vec{r}'} V_{\vec r\vec r'}(\phi_0) Q\svr Q_{\vec r'} - \sum\svr h\svr Q\svr,
\end{align}
where $Q\svr$ is a JT (breathing) distortion for the manganites (nickelates) and  
$\Pi\svr$ its conjugate momentum.  To model the compositional disorder that arises in the manganites from chemical substitution of the 
alkaline earth element at the $A$ site of the perovskite structure,
we consider a linear coupling of the lattice distortions $Q\svr$
to a local quenched random distortion $h\svr$.
We choose the $h\svr$'s to be normally distributed with mean $\bar h\svr=0$ and variance $\bar h^2\svr = \Delta^2$. The negative sign of the $Q\svr^2$ term describes the local tendency towards distortion due to the presence of electrons.

\begin{figure*}[htp]
  \centering
  \includegraphics[scale=0.310]{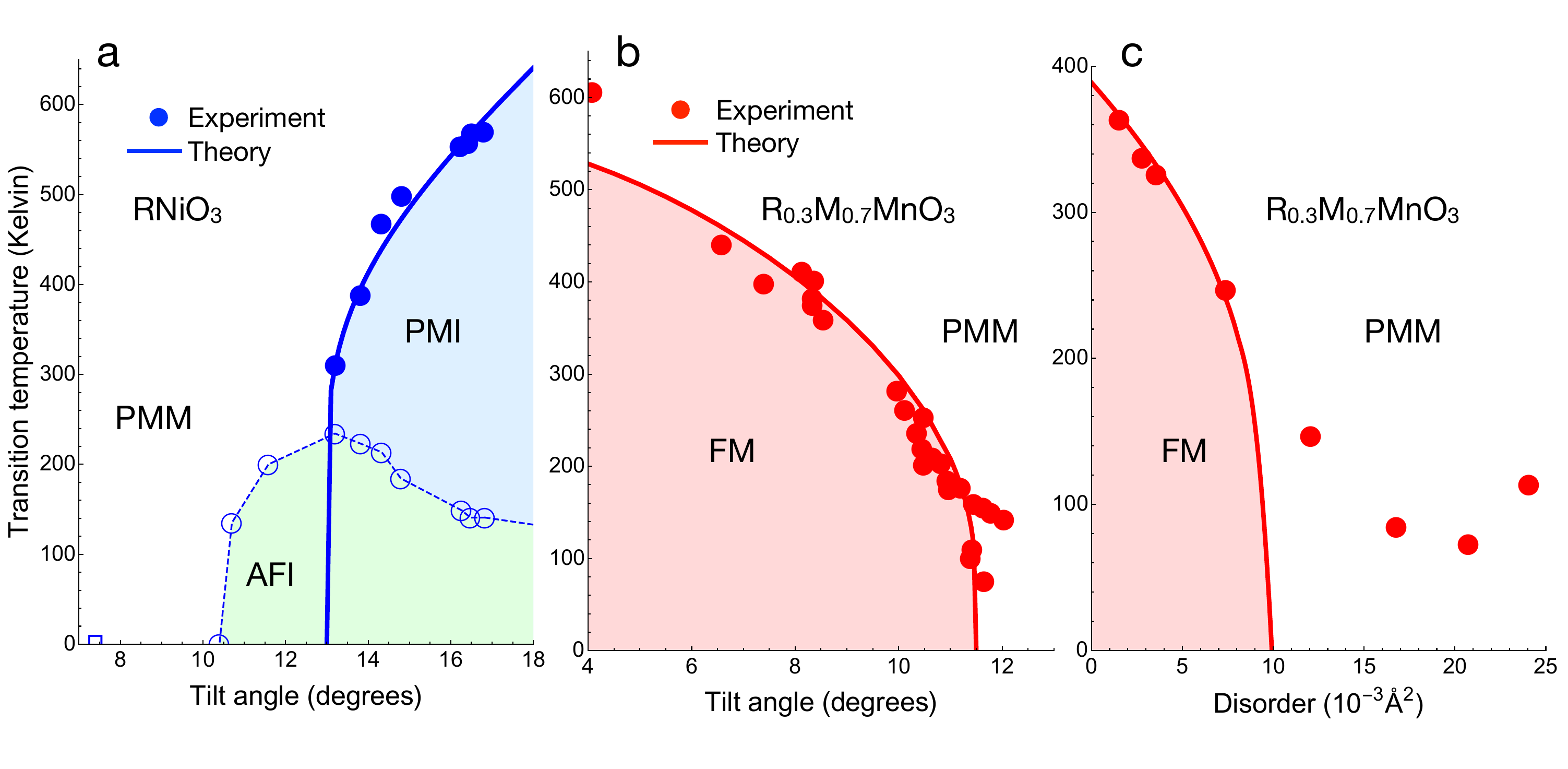}
  \caption{ {\bf Comparison to experiments.}
  {\bf (a)} Comparison for the nickelates for the transition temperature as a function of octahedral tilt angle. Filled and open circles are experimental transition temperatures~\cite{Catalano2018a} for the paramagnetic insulator (PMI) and antiferromagnetic insulator (AFI) phases respectively. The extension of the green shading beyond the blue dashed line is an extrapolation. The open square denotes $\mathrm{LaNiO}_3$, which is always in the high temperature polaronic, paramagnetic metal (PMM) phase. For the manganites, the comparison is made with results~\cite{RodriguezMartinez1996a} (red circles) that separate the effect of tilt angle {\bf (b)} and compositional disorder {\bf (c)} on the transition from the paramagnetic,  polaronic bad metal phase (PMM) to the ferromagnetic metal (FM) phase.  Model parameters are fitted as described in Methods. In Supplementary Note 3, we explore reasonable variations of the model parameters to demostrate its generality.}
  \label{fig:phase-diagrams}
\end{figure*}

As described in Methods and Supplementary Note 2,  we use a variational approach 
to  calculate the temperature, tilt angle and disorder dependence of the  free energy $F_\text{lattice}\left(T, \phi_0, \Delta \right)$ of Hamiltonian (\ref{eq:Hamiltonian}); and 
we identify the location of $T_\text{MI}$  by comparing $F_\text{lattice}\left(T, \phi_0, \Delta \right)$ to a  free energy $F_{\text{low}\, T}$ of the low temperature FM (PMI) phase of the manganites (nickelates).
The results are shown in Fig.~\ref{fig:phase-diagrams}. Despite the over-simplicity of the model, the relationship between tilt angle, disorder, and transition temperature is well reproduced.  We do not attempt to describe the effects of the strain interactions on the MIT of the nickelates at low temperatures (see green region in  Fig.~\ref{fig:phase-diagrams}\,(a)), as its magnetic ordering  is different from that of the insulating phase above it. Similarly for the manganites, at low enough temperatures the PMM phase becomes either charge-ordered or glassy, beyond our approximations.
In this paper we have outlined a systematic theory for the incorporation of long-range elastic couplings into a simplified statistical mechanical theory of Mott-like phase transitions, where the electronic contributions to the free energy are incorporated at the level of Landau theory. That these elastic interactions are explicitly relevant for the manganites and the nickelates is confirmed by the ability of such a theory to systematically explain ‘size effects’ or ‘tolerance factor’ variations which have already been documented. However the couplings, including their rough order-of-magnitude, are generic, and the ideas presented here will surely be relevant to other classes of materials such as the titanates~\cite{Katsufuji1997a}, high temperature superconductors~\cite{Attfield1998a}, ferroelectrics~\cite{Balachandran2011a}, and molecular fullerides~\cite{Zadik2015a}. 

At low enough temperatures one should surely take care of other low-energy degrees of freedom such as spin fluctuations and electronic quantum fluctuations which our model does not take into account. Doing so requires explicitly adding them to our model Hamiltonian and to our statistical mechanical solution through, e.g. a variational scheme such as the Lang-Firsov transformation.
Nonetheless, the model we employ does generate a quantum critical point on account of elastic interactions alone. Moreover, the long range and anisotropy of these elastic couplings will  modify the critical dynamics away from that arising from short-range models generated by purely electronic couplings.

We also note that our simple model provides an explanation for the observed tuning of the MIT 
 under applied pressures.
In both the manganites~\cite{Radaelli1997a} and nickelates~\cite{Obradors1993a},
hydrostatic compression decreases $\phi_0$. According to our model, 
this should result in an increase of $T_\text{MI}$ promoted by the enhancement 
of the elastic interaction in the manganites, and viceversa for the nickelates. 
These are indeed the trends that have been observed in these materials.~\cite{Fontcuberta1998a, Obradors1993a}  
We believe a similar mechanism is at play when the transition is tuned with tensile and compressive stresses~\cite{Liu2013a}.
 
The idea that cooperative phonon-phonon couplings tune the MIT 
is supported by a recent ab-initio calculation~\cite{Mercy2017a}. 
By using density functional theory~(DFT),  it has been found that the tilts of the NiO$_6$ units in the nickelates destabilize their breathing distortions, which in turn
are associated with the phase transition, thus providing  a mechanism for tuning $T_\text{MI}$. 
However, DFT treats the elastic interactions only in average and it cannot produce finite temperature properties,
thus $T_\text{MI}$ was obtained by fitting it to experiments  with
a Landau theory that has multiple sets of values for the model parameters depending on the tolerance factor. 
By contrast, we have calculated $T_\text{MI}$ from a single set of model parameters, and the MIT is driven 
by entropic effects that result from elastic couplings, thus providing a physical interpretation of the ab-initio results.

We conclude by noting that the good agreement we found in these two systems suggests that our fundamental assumption that the energy could be separated into a relatively simple local free energy plus a complex long range piece, could provide a basis for a fully computational methodology that could be applied relatively simply to very complex oxides in general. 

\newpage

\noindent {\bf Acknowledgements.} We acknowledge insightful discussions with G. Lonzarich, H. Park and F. Ballar-Trigueros.
Work at Argonne National Laboratory is
supported by the U.S. Department of Energy, Materials Science Division, Office of
Basic Energy Sciences under contract \linebreak
no. DE-AC02-06CH11357.
G.G.G.-V. acknowledges  support from the
Vice-rectory for Research (project  no. 816-B7-601), and the 
Office of International Affairs at the University of Costa Rica,  
the Royal Society International Exchanges programme (IES\textbackslash R3\textbackslash 170025), 
Churchill College (Cambridge),
and thanks the Department of Materials Science and Metallurgy and 
the Cavendish Laboratory at the University of Cambridge for hospitality where part of this work was done. R.T.B. acknowledges  support from the Yale Prize Postdoctoral Fellowship and Homerton College (Cambridge).

\vspace{0.25cm}

\noindent {\bf Methods}

\vspace{0.25cm}

 \noindent {\bf Statistical mechanical solution.} 
We use a variational pair-distribution function that incorporates mean-field behavior, Gaussian corrections to the thermal and quantum fluctuations, and averaging over compositional disorder at the level of the replica method~\cite{GuzmanVerri2013a}. Details are provided in Supplementary Note 2.

 \noindent {\bf Model parameters.}   
Our model has six parameters ($\kappa, \gamma, a_D, a_S, a_T,$ and $ F_{\text{low}\, T} $), which are reduced to five as $a_T$ ($a_D$) is combined with $\kappa$ for the manganites (nickelates).
We begin by choosing  a set of  of physically reasonable parameters which give phonon frequencies that are in order-of-magnitude agreement with 
the observed relevant modes~\cite{Zaghrioui2001a, Martin2002a}. We then take the resulting set of parameters and fine tune them 
to fit the observed dependence of $T_\text{MI}$ with the tolerance factor and compositional disorder:
 $F_{\text{low}\, T}$ is a parameter of the model assumed to be independent of $T, \phi_0$, and $\Delta$, fixed by the observed onset of the MIT, i.e., $F_\text{lattice}\left(T=0\,\text{K}, \phi_0 = \phi_\text{onset}, \Delta =0 \right) = F_{\text{low}\, T}$, where $\phi_\text{onset} \simeq 11.5^\circ\,$($12.5^\circ$) for the manganites (nickelates).  The dependence of $T_\text{MI}$ on $\phi_0$ shown in Figs.~\ref{fig:phase-diagrams}\,(a) and \ref{fig:phase-diagrams}\,(b) is given by  $F_\text{lattice}\left(T_\text{MI}, \phi_0, \Delta =0 \right) = F_{\text{low}\, T}$, while 
 the dependence of $T_\text{MI}$ on $\Delta$ shown in Fig.~\ref{fig:phase-diagrams}\,(c)  is given by  $F_\text{lattice}\left(T_\text{MI}, \phi_0=8.0^\circ, \Delta \right) = F_{\text{low}\, T}$ and by rescaling $\Delta$ by a constant factor ($\alpha$) to match the units of cation variance.  The resulting values  are given in Table~\ref{t:parameters}.

\newpage

\begin{table}[htp]
   \caption{Model parameters.
                 R is a rare earth element such as  La, Pr, Nd, and Sm;  and M is an alkaline earth metal such as Ca, Sr, Ba.}
\begin{adjustbox}{angle=90}
   \begin{tabular}{crrcrccc} \hline
                            & $   \kappa\,$ [meV$^2$]  & $\gamma\,$ [meV$^3$] & $a_D\,$ [meV$^2$]  & $ a_S\,$ [meV$^2$]                                  & $ a_T\,$ [meV$^2$]  & $ F_{\text{low}\, T}$ [meV]  & $\alpha$ [\AA\,meV$^{-1}$]    \\ \hline \hline
     (R$_{0.3}$M$_{0.7}$)MnO$_3$  &          $3.2 \times 10^3$ &       $1.8 \times 10^5$ &     $ 2.8 \times 10^4 $          &     $ 9.5 \times 10^3$  &  combined with $\kappa$   &    $38$  & 4.71   \\
           RNiO$_3$                         &         $12.1 \times 10^3  $   &	 $13.3 \times 10^5$&	  combined with $\kappa$        &  $48.3\times10^3$   & $1.21 \times10^{4}$ &  $64 $ & - \\ \hline
   \end{tabular}
   \end{adjustbox}
   \label{t:parameters}
\end{table} 

\newpage


%

\clearpage
\pagebreak
\widetext
	\begin{center}
		\textbf{\large Supplementary Information \\ for \\ Cooperative elastic fluctuations provide tuning of the metal-insulator transition}
	\end{center}
\vspace{0.5cm}
	\begin{center}
		G. G. Guzm\'{a}n-Verri$^{1,2,3}$, R. T. Brierley$^4$, P. B. Littlewood$^{3,5}$ \\
		$^{1}${\it Centro de Investigaci\'{o}n en Ciencia e Ingenier\'{i}a de Materiales~(CICIMA),\\ Universidad de Costa Rica, San Jos\'{e}, Costa Rica 11501,}\\
		$^{2}${\it Escuela de F\'{i}sica, Universidad de Costa Rica, San Jos\'{e}, Costa Rica 11501,} \\
		$^{3}${\it Materials Science Division, Argonne National Laboratory, Argonne, Illinois, USA 60439,}\\
		$^{4}${\it Department of Physics, Yale University, New Haven, Connecticut 06511, USA,}\\
		$^{5}${\it James Franck Institute, University of Chicago, 929 E 57 St, Chicago, Illinois, USA 60637.}
		\end{center}

\vspace{0.5cm}
\begin{center}
	\textbf{Contents} \\
	\begin{itemize}
		\item[] {\textbf{Note 1. Distortion interaction.}} 
		\item[] {\textbf{Note 2. Variational Solution.}}
		\item[] {\textbf{Note 3. Model parameter dependence of the phase diagrams.}}
		\item[] {\textbf{References.}}	
	\end{itemize}
\end{center}

\setcounter{equation}{0}
\setcounter{figure}{0}
\setcounter{table}{0}
\setcounter{page}{1}
\makeatletter
\renewcommand{\theequation}{S\arabic{equation}}
\renewcommand{\bibnumfmt}[1]{[S#1]}
\renewcommand{\citenumfont}[1]{S#1}

\newpage

\section*{\label{sec:SN1} Supplementary Note 1. Distortion interaction}
\begin{figure}[h!] 
     \centering
  \includegraphics[scale=0.25]{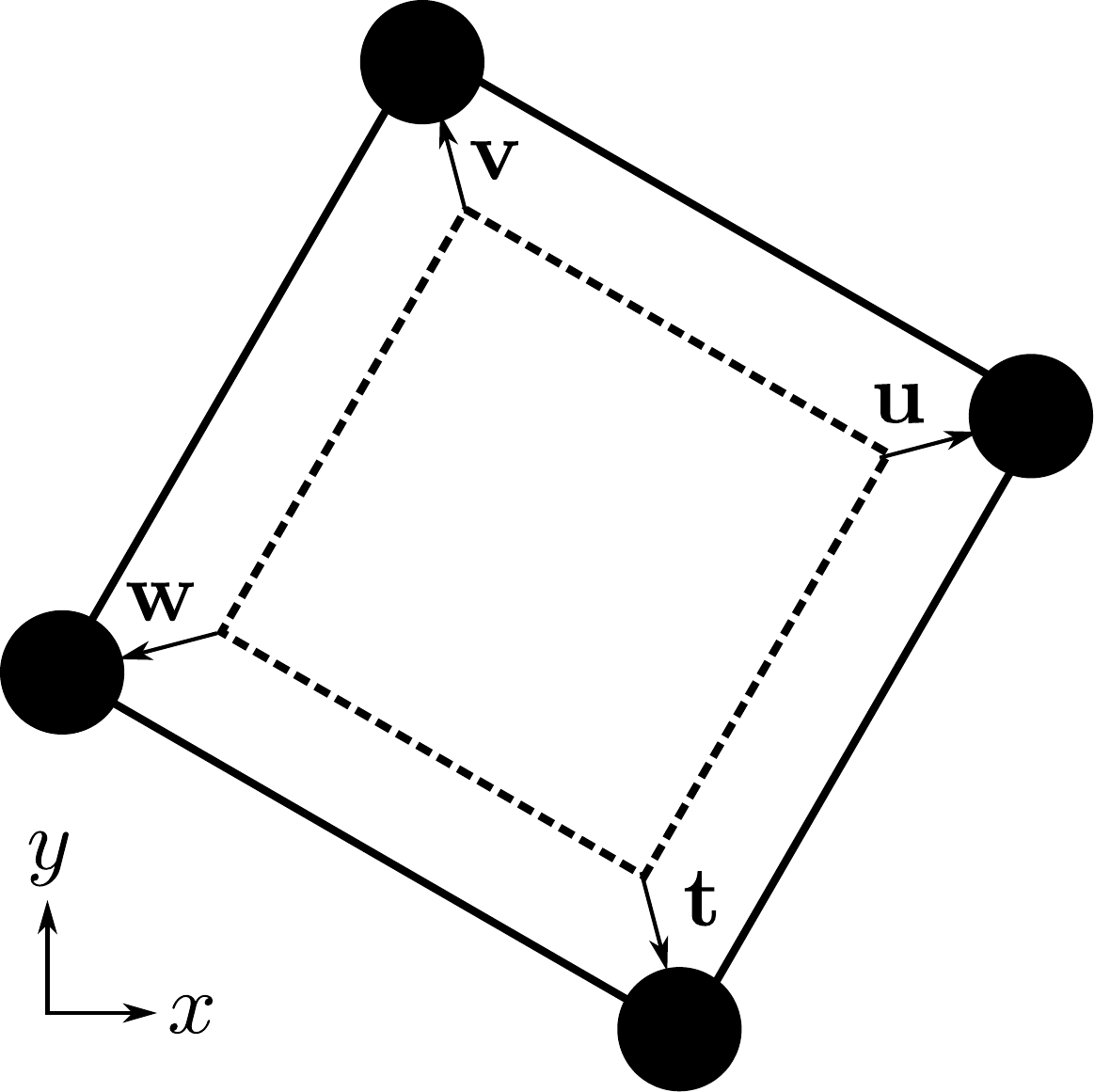}
   \caption{ {\bf Quadrilateral displacements.}  A quadrilaterial altered from its equilibrium state (dashed) by a breathing distortion. The corner atoms displaced by displacement vectors~$\vec t\svr$,~$\vec u\svr$,~$\vec v\svr$~and~$\vec w\svr$.}
  \label{fig:distortions}
\end{figure}

We consider a two-dimensional model of a perovskite, consisting of corner-coupled quadrilaterals that are the analogues of the three-dimensional oxygen octahedra. The displacements of the corners of the quadrilateral centred at a position $\vec r$ from their initial positions are the vectors $\vec t\svr$, $\vec u\svr$, $\vec v\svr$ and $\vec w\svr$ (Supplementary Fig.~\ref{fig:distortions}). To simplify the analysis, we assume that the allowed (i.e.~low energy) configurations of the quadrilaterals are parallelograms, so that
\begin{gather}
  \vec t\svr-\vec w\svr = \vec u\svr-\vec v\svr\label{eq:parallelogram-cond}.
\end{gather}
This corresponds to assuming there is infinite energy cost associated with ``shuffle'' modes within the a quadrilateral \cite{Sahn_atomic_2003}. With this assumption, deviations from an initial configuration can be described in terms of four degrees of freedom: a rotation and three ``strains'': $D\svr$ (dilatation/breathing modes), $S\svr$ (shear modes) and $T\svr$ (deviatoric/JT modes).
\captionsetup[figure]{labelfont=bf,name=Supplementary Fig.,labelsep=period, justification=justified}
We take the following simple form for the elastic energy,
\begin{gather}
  \label{Seq:energy-unsubbed}
  H = \sum\svr a_TT\svr^2 + a_DD\svr^2 + a_SS\svr^2,
\end{gather}
where $a_i$ is the energy penalty for the corresponding distortion of a free octahedron.
Note that we make the approximation that the cost of rotations is small enough that it can be negelected.
Such a term should exist in order for there to be a state with a non-zero $\phi_0$, as observed in experiment.
In the perovskite model, the constraint that the squares share corners means that $\vec t\svr = \vec v\svdispy$ and $\vec w\svr = \vec u\svdispx$ where $\vec n_i$ are the lattice vectors. 
These constraints reduce the number of degrees of freedom per site to 2.
Distortions of neighboring squares are coupled and as a result the energy given in 
Eq. (\ref{Seq:energy-unsubbed}) actually describes a system with long-range strain interactions.

We can find the constraint equations in terms of the rotation and strain degrees of freedom by expressing them in terms of the atom positions $\vec t\svr$, $\vec u\svr$, $\vec v\svr$ and $\vec w\svr$, then using the corner-sharing constraint to eliminate the positions.
We assume that rotations of the squares from an initial equilibrium antiferrodistortive rotation $\phi_0$ are small, writing $\phi\svr = e^{-i\vec{M}\cdot\vec{r}}\phi_0 + R\svr$, where $\vec{M}=(\pi/L,\pi/L)$ and $L$ is the equilibrium lattice spacing.
Then, linearising in the small rotations $R\svr$, we can write:
\begin{subequations}
\label{eq:strain-and-displacement-relationship}
\begin{gather}
  \label{eq:strain-and-displacement-relationship-rr}
  R\svr = \cos\phi_0(u^y\svr-u^y\svdispx-v^x\svr+v^x\svdispy) - e^{-i\vec M \cdot \vec r}\sin\phi_0(u^x\svr-u^x\svdispx+v^y\svr-v^y\svdispy),\\
        \label{eq:strain-and-displacement-relationship-sr}
      S\svr = \cos\phi_0(u^y\svr-u^y\svdispx+v^x\svr-v^x\svdispy) - e^{-i\vec M \cdot \vec r}\sin\phi_0(u^x\svr-u^x\svdispx-v^y\svr+v^y\svdispy),\\
    \label{eq:strain-and-displacement-relationship-tr}
    T\svr = \cos\phi_0(u^x\svr-u^x\svdispx-v^y\svr+v^y\svdispy) + e^{-i\vec M \cdot \vec r}\sin\phi_0(u^y\svr-u^y\svdispx+v^x\svr-v^x\svdispy),\\
        \label{eq:strain-and-displacement-relationship-dr}
  D\svr = \cos\phi_0(u^x\svr-u^x\svdispx+v^y\svr-v^y\svdispy) + e^{-i\vec M \cdot \vec r}\sin\phi_0(u^y\svr-u^y\svdispx-v^x\svr+v^x\svdispy),
\end{gather}
\end{subequations}
where we have used the convention $\vec u\svr = (u^x\svr,u^y\svr)$ for the vector components.

After Fourier transformation, the parallelogram condition Eq.~\eqref{eq:parallelogram-cond}, combined with the corner sharing constraint, can be written as,
\begin{gather}
  \label{eq:ft-parallelogram}
  (1+e^{ik_xL})\vec u \svk = (1+e^{ik_yL})\vec v\svk.
\end{gather}
Using this relationship, the Fourier transform of Eq.~\eqref{eq:strain-and-displacement-relationship} is,
\begin{gather}
  \label{eq:strain-and-displacement-ft}
  \begin{aligned}
  R\svk = \left(1+e^{ik_xL}  \right)\left[ i \cos\phi_0 \left(t_{k_x} u^y\svk-t_{k_y} u^x\svk\right) -\frac{\sin\phi_0}{t_{k_y}}\left(t_{k_y} u^x\svpik+t_{k_x} u^y\svpik  \right) \right]\text,\\
  S\svk = \left(1+e^{ik_xL}  \right)\left[i \cos\phi_0 \left(t_{k_x} u^y\svk+t_{k_y} u^x\svk\right) -\frac{\sin\phi_0}{t_{k_y}}\left(t_{k_y} u^x\svpik -t_{k_x} u^y\svpik \right)\right]\text,\\
  T\svk = \left(1+e^{ik_xL}  \right)\left[i \cos\phi_0 \left( t_{k_x} u^x\svk -t_{k_y} u^y\svk\right) +\frac{\sin\phi_0}{t_{k_y}}\left(t_{k_y} u^y\svpik + t_{k_x} u^x\svpik \right)\right]\text,\\
  D\svk = \left(1+e^{ik_xL}  \right)\left[i \cos\phi_0 \left( t_{k_x} u^x\svk + t_{k_y} u^y\svk\right) +\frac{\sin\phi_0}{t_{k_y}}\left(t_{k_y} u^y\svpik - t_{k_x} u^x\svpik \right)\right]\text,
  \end{aligned}
\end{gather}
where we have defined,
\begin{gather}
  \label{eq:ti-definition}
  t_{k_i} = \tan\frac{k_iL}2\text.
\end{gather}
Then, using the fact that that $t_{k_i}\to -1/t_{k_i}$ when $\vec k \to \vec k-\vec M$ we can write: 
\begin{gather}
    \begin{aligned}
    u^x\svk&=\frac{i e^{-\frac{ik_xL}{2}}\left[\ty \sin \phi_0(T\svpik-D\svpik)-\tx\cos\phi_0(T\svk+D\svk)\right]}
    {4\cos\frac{k_xL}2  \left(\tx^2\cos^2\phi_0+\ty^2\sin^2\phi_0 \right) }\text,\\
    u^y\svk&=\frac{-ie^{-\frac{ik_xL}{2}}\left[\ty\cos\phi_0(D\svk-T\svk)-\tx\sin\phi_0(T\svpik+D\svpik)\right]}
    {4\cos\frac{k_xL}{2}\left(\ty^2\cos^2\phi_0+\tx^2\sin^2\phi_0\right)}\text.
    \end{aligned}
\end{gather}$\vec u\svk$ can now be eliminated from Eq.~\eqref{eq:strain-and-displacement-ft}. We obtain compatibility conditions between the distortions:
\begin{gather}
    \label{eq:discrete-compat-rk}
    \begin{aligned}
      R\svk = f\svk\Big\{
      &4\tprod\left[ -\tplus T\svk + \cos2\phi_0\tmin D\svk\right]\\ +
      &2\sin2\phi_0\tmin\left[ \tplus T\svpik + \cos2\phi_0\tmin D\svpik \right]
      \Big\}\text,
\end{aligned}\\
          \label{eq:discrete-compat-sk}
  \begin{aligned}
    S\svk = f\svk \Big\{
      &4\tprod\left[ \tplus D\svk - \cos2\phi_0\tmin T\svk\right]\\ -
      &2\sin2\phi_0\tmin\left[\tplus D\svpik +\cos2\phi_0\tmin
        T\svpik \right] \Big\}\text.
    \end{aligned}
\end{gather}
where
\begin{gather}
  \label{eq:denom-func}
  f\svk = \frac{1}{8 \tprodsq+ 2\tmin^2 \sin^22\phi_0}\text.
\end{gather}
In later manipulations it is useful to use the relationship $f\svpik=\tx^4\ty^4 f\svk$.

These relationships have singular behavior when the $k_i$ approach $0$ or $\pi$.
In particular, there is no relationship between the fields when $\vec k=\vec 0$ or $\vec M$, i.e.~the cases of homogeneous and antiferrodistortive deformations; each strain field is a separate degree of freedom in these cases \cite{Slarkin_phase_1969}. In the long-wavelength limit, $t_{k_x}\to  k_xL /2$ and when $\phi_0=0$ we can write Eq.~\eqref{eq:discrete-compat-sk} as,
\begin{gather*}
  \label{eq:norot-sk}
  (k_x^2+k_y^2)D\svk -2 k_x k_y S\svk - (k_x^2-k_y^2)T\svk = 0,
\end{gather*}
which is the Fourier transform of the usual two-dimensional compatibility relation\cite{SKartha1995a}.

In preparation to find $H$ in terms of $D\svk$ and $T\svk$ we now evaluate:
\begin{gather}
\begin{aligned}
    \sum\svk |S\svk|^2& =\sum\svk f\svk^2 \left\{16 \tprodsq\tplus^2 |D\svk|^2 + 4\sin^22\phi_0\tminfour^2|D\svpik|^2\right.\\ &+16\tprodsq\tmin^2\cos^22\phi_0|T\svk|^2+4\sin^22\phi_0\cos^22\phi_0\tmin^4|T\svpik|^2\\
    &-16\tprodsq\tminfour\cos2\phi_0(D^*\svk T\svk+\text{h.c.})\\
    &+4\sin^22\phi_0\cos2\phi_0\tmin^2\tminfour(D^*\svpik T\svpik+\text{h.c.})\\
    &-8\sin2\phi_0\tprod\tmin\left[\tplus^2 D^*\svk D\svpik-\cos^22\phi_0\tmin^2T^*\svk T\svpik\right.\\
    &+\left.\left.\cos2\phi_0\tminfour( D^*\svk T\svpik-T^*\svk D\svpik)+\text{h.c}\right]
    \right\}\text.
\end{aligned}
\end{gather}
By considering both the $\vec k$ and $\vec k-\vec M$ parts of the sum the terms that couple fields at $\vec k$ and $\vec k - \vec M$ cancel and the rest of the expression can be simplified to:
\begin{gather}
    \begin{aligned}
    \sum\svk|S\svk|^2&=\sum\svk f\svk\left[2\tplus^2|D\svk|^2+2\cos^22\phi_0\tmin^2|T\svk|^2\right.\\
    &\left.-2\cos2\phi_0\tminfour(D^*\svk T\svk+\text{h.c.})\right]\text.
    \end{aligned}
\end{gather}
Substituting this into the Fourier transform of Eq.~\eqref{Seq:energy-unsubbed} we get the result without separate constraints:
\begin{gather}
  \label{eq:energy-subbed}
  \begin{split}
    H=\sum_{\vec{k}=\vec 0}^{(\pi,\pi)} \left[ 
      a_T+2a_Sf\svk\tmin^2\cos^22\phi_0
     \right]|T\svk|^2 +\left[  a_D + 2a_S f\svk\tplus^2
    \right] |D\svk|^2 \\-2a_Sf\svk\tminfour\cos2\phi_0(D\svk^*T\svk+\text{h.c.})\text.
  \end{split}
\end{gather}

To obtain the static interaction between the single strain fields $T\svk$ or $D\svk$, we can minimize the energy with respect to the other field:
\begin{gather}
  \label{eq:minimal-tk-or-dk}
  D\svk = \frac{2a_Sf\svk\tminfour\cos2\phi_0}{a_D+2a_Sf\svk\tplus^2}T\svk,\quad \text{or}\quad T\svk = \frac{2a_Sf\svk\tminfour\cos2\phi_0}{a_T+2a_Sf\svk\tmin^2\cos^22\phi_0}D\svk. 
\end{gather}
Substituting these results into Eq.~\eqref{eq:energy-subbed}, we obtain the  energy for a single strain field, including the long-range interaction term,\\
 \label{eq:free-energy}
\begin{align}
  \label{eq:free-energy-tk}
  H_T &= \sum\svk\left\{ a_T + \frac{a_Da_S\cos^22\phi_0\tmin^2}{4a_D\tprodsq+a_S\tplus^2+a_D\tmin^2\sin^22\phi_0} \right\} |T\svk|^2 \\
  \label{eq:free-energy-dk}
  H_D &=   \sum\svk\left\{ a_D + \frac{a_S a_T\tplus^2}{4 a_T\tprodsq + a_S\tmin^2 + (a_T-a_S)\tmin^2\sin^22\phi_0} \right\} |D\svk|^2.
\end{align}
These strain interactions are highly anisotropic~(see Fig. 3 in manuscript)
and non-analytic; as $|\vec k|\to 0$, the value of the potential depends on the direction of $\vec k$. 
The potential vanishes for (anti-)ferrodistortive perturbations, corresponding to $\vec k =0$ ($\vec k=\vec M = (\pi,\pi)$), as the different distortion modes are uncoupled at those wavevectors. This implies a discontinuity at $\vec k=\vec M$, which arises from our assumption that 
the distortions in Fig.~\ref{fig:distortions} are the only distortions, neglecting the shuffle modes for the octahedra.\cite{Sahn_atomic_2003} However, as long as such modes are significantly stiffer than the high-symmetry modes we do consider, this should not be an important approximation.
In the long-wavelength limit  and no rotations,
$ V_{\vec k}(\phi_0=0)  $ for the manganites matches  that of previous work.~\cite{SKartha1995a, SPorta2009a}

\newpage

\section*{\label{sec:SN2} Supplementary Note 2. Variational Solution}

We consider a trial pair-probability distribution,
$\rho^{tr} = (Z^{tr})^{-1} e^{-\beta H^{tr}},$
where $H^{tr}$ is the Hamiltonian of  coupled harmonic oscillators in a random field,
$H^{tr} = \sum\svr  \frac{1}{2} \Pi\svr^2 + \frac{1}{2} \sum_{\vec{r} \vec{r}'} Q\svr \mathcal{D}_{\vec r - \vec r'}   Q\svr^\prime  - \sum_{\vec r} h\svr Q\svr$
and $Z^{tr} = \mbox{Tr} e^{- \beta H^{tr} }$ its normalization. 
The Fourier transform of the function $\mathcal{D}_{\vec r - \vec r'}$ gives the  frequency squared of the mode at wavevector ${\vec k}$,
i.e.,  $ \Omega_{\vec k}^2 =  \sum_{\vec{r} \vec{r}'} \mathcal{D}_{\vec r\vec r'}  e^{ i {\vec k} \cdot (\vec r - \vec r') } $.
  $\Omega_{\vec k}^2$  is a variational function and it is determined by minimization of the free energy
of the lattice degrees of freedom, $ F_\text{lattice} =   \overline{ \bra H \ket } + T  \overline{ \bra k_B \ln \rho^{tr} \ket } $.
Here, $  \overline{ \bra .... \ket } $ denotes thermal and compositional averages. 
The  free energy per site is therefore given by,
\begin{align}
\label{eq:Sfree}
\frac{ F_\text{lattice} }{N} &= -\frac{\kappa}{2} \left( \frac{1}{N} \sum_{\vec k}  \eta_{\vec k} \right) 
 + \frac{3 \gamma}{4} \left( \frac{1}{N} \sum_{\vec k}  \eta_{\vec k} \right)^2 
 + \frac{1}{2N} \sum_{\vec k} V_{\vec k}(\phi_0) \, \eta_{\vec k} \\
 &~~~~~~~~~~~~
 -  \frac{1}{N} \sum_{\vec k} \frac{\Delta^2}{ \Omega_{\vec k}^2}
- \frac{1}{4N} \sum_{\vec k}  \Omega_{\vec k} \coth \left( \frac{ \beta  \Omega_{\vec k} }{2} \right)  
+ \frac{k_B T}{N} \sum_{\vec k}  \ln \left[ 2 \sinh\left( \frac{ \beta  \Omega_{\vec k} }{2} \right)  \right], \nonumber
\end{align}
where $ \eta_{\vec k} = \frac{1}{2  \Omega_{\vec k}} \coth \left( \frac{ \beta  \Omega_{\vec k} }{2} \right) + \frac{\Delta^2}{ \Omega_{\vec k}^4 }$
are mean squared fluctuations averaged over compositional disorder at wavevector ${\vec k}$.  $N$ is the number of lattice sites and the summations run over the 
first Brillouin zone of the square lattice. Minimization of the free energy (\ref{eq:Sfree}) with respect to $\Omega_{\vec k}$ gives the following result, 
\begin{align}
\label{eq:Seuler-lagrange}
 \Omega_{\vec k}^2 &=  - \kappa + 3 \gamma \, \frac{1}{N}\sum_{\vec k^\prime} \left[ \frac{1}{2  \Omega_{\vec k^\prime}} \coth \left( \frac{ \beta  \Omega_{\vec k^\prime} }{2} \right) + \frac{\Delta^2}{ \Omega_{\vec k^\prime}^4 } 
 \right] + V_{\vec k}. 
\end{align}
Equation (\ref{eq:Seuler-lagrange}) determines the temperature and disorder dependence of the mode frequency $\Omega_{\vec k}$ self-consistently.

\newpage

\section*{\label{sec:SN3} Supplementary Note 3. Model parameter dependence of the phase diagrams}

\begin{figure}[h!]
  \centering
    \includegraphics[scale=0.5]{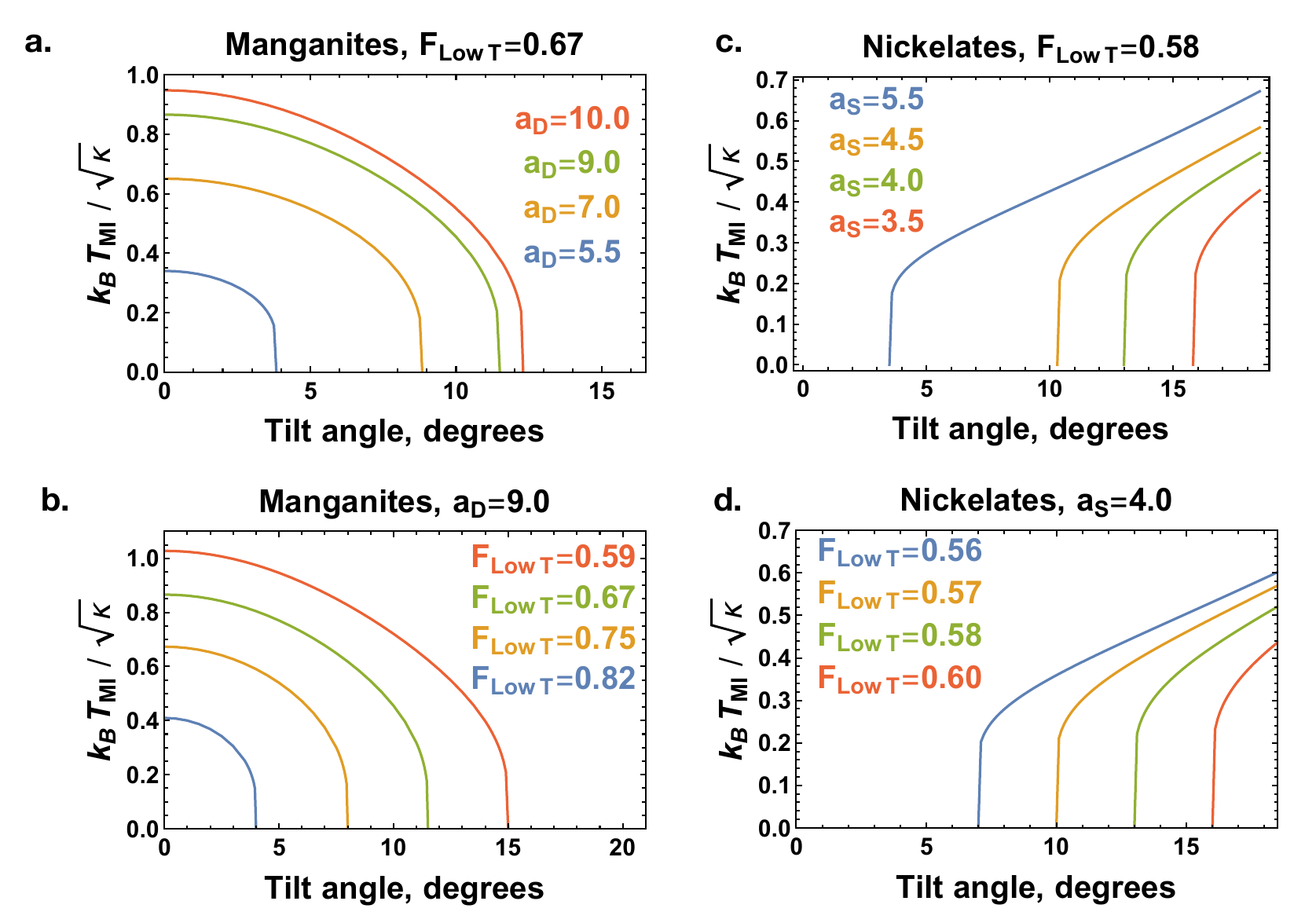}
    \caption{ {\bf Phase diagrams.}   Calculated phase diagrams for the {\bf (a)-(b)} manganites  and {\bf (c)-(d)} nickelates  for several breathing ($a_D$) and shear ($a_S$) stiffness and energies of their corresponding low-temperature magnetic phases ($F_{\text{low}\,T}$). For (a)-(b), $a_S=3.0$ and for (c)-(d), $a_T=1.0$ . $\gamma=1$  for all plots. $F_{\text{low}\,T}, a_{D/S/T}$, and $\gamma$ are, respectively in units of $\kappa^{1/2}, \kappa$,  and, $\kappa^{3/2}$.}
  \label{fig:phase_diagrams}
\end{figure}

Supplementary Fig.~\ref{fig:phase_diagrams} shows how the calculated phase diagrams vary with the relevant model parameters. By changing the values of those parameters so as to keep the transition temperatures physical, we find that the trends in  $T_\text{MI}$ remain, 
while the quantitative agreement with experiments decreases 
(green curves correspond to those of Fig.~4.).  
These results also show that we cannot choose parameters that would make the elastic effects on the MIT small. Only a widely different parametrization would lead to different behavior and destroy the  agreement.

\captionsetup[figure]{labelfont=bf,name=Supplementary Fig.,labelsep=period}

\end{document}